\documentclass [reprint,pra,aps,showpacs]{revtex4-1}
\usepackage{amssymb, amsthm}
\usepackage{graphicx}
\usepackage[dvips]{color}
\usepackage{multirow}

\def\eps{\varepsilon}
\def\bE{{\mathbb E}}
\def\Var{\mathrm{Var}\,}
\def\Cov{\mathrm{Cov}\,}

\begin{document} 

\title{Long-distance continuous-variable quantum key distribution with efficient channel estimation}
\author{L\'aszl\'o Ruppert$^{1}$
, Vladyslav C. Usenko$^{1,2}$ and Radim Filip$^1$}
\affiliation{$^1$ Department of Optics, Palacky University, 17. listopadu 12, 771 46 Olomouc, Czech
Republic\\
$^2$ Bogolyubov Institute for Theoretical Physics of National Academy of Sciences,
Metrolohichna st. 14-b, 03680, Kiev, Ukraine}


\begin{abstract}
We investigate the main limitations which prevent the continuous-variable quantum key distribution protocols from achieving long distances in the finite-size setting. We propose a double-modulation protocol which allows using each state for both channel estimation and key distribution. As opposed to the standard method, we optimize the parameters of the protocol and consider squeezed as well as coherent states as a signal. By optimally combining the resources the key rate can approach the theoretical limit for long distances, and one can obtain about ten times higher key rate using ten times shorter block size than in the current state-of-the-art implementation.
\end{abstract}

\maketitle


\section{Introduction} 
The rapid development of experimental quantum optics in the past decades has led to the establishment of quantum key distribution (QKD) \cite{Gisin02,Scarani09}. This branch of quantum information science is aimed at the development and implementation of methods for the secure distribution of cryptographic keys so that the very laws of quantum physics provide the security of the key. The key can be then used in the one-time-pad classical cryptographic system, thus providing the possibility ideally for an unconditionally secure information transmission. 

A recent addition to QKD is the use of homodyne detection of continuous-variable (CV) states \cite{Weed2012}. Enforced by the Gaussian security proofs \cite{wolf06,Nav06,GP06,Lev13}, this allows the use of multiparticle quantum states in the typical photonic implementations of QKD in order to improve the applicability and stability of QKD protocols. CV QKD protocols were first suggested and studied on the basis of nonclassical squeezed states of light \cite{Ralph99,Cerf01}. An important step in CV QKD was made when the semiclassical coherent states were shown to be in principle sufficient for key distribution \cite{Grosshans02} even at long distances, using Gaussian modulation and reverse information reconciliation \cite{Grosshans03}. The coherent-state protocol was successfully tested in mid-range (25 km) \cite{Lod10} and long-range (80 km) \cite{Jou13} optical fiber channels, while a proof-of-principle laboratory test of the squeezed-state protocol was performed recently \cite{Mad12}. It was shown in particular that the squeezed-state protocol can potentially outperform its coherent-state counterpart in terms of robustness and range, especially if data processing efficiency is limited \cite{Use11}. 

For the security analysis of the CV QKD protocols the trusted parties need to know the correlation of their measurements. Because of the optimality of Gaussian states, that is equivalent to knowing the parameters of the channel, namely, its transmittance and excess noise. This knowledge allows one to estimate the maximum amount of information leaking to a potential eavesdropper from such a channel. The channel estimation is, however, a nontrivial task. Indeed, the trusted parties need to estimate the channel based on probe pulses, which must be indistinguishable from the signal, otherwise the eavesdropper can recognize them and exploit that information. The accuracy of the estimation is typically limited by the intensity and number of estimation pulses;  however, more states used for estimation means a lower capacity to carry information. The effect of the limited ensemble size on the applicability of CV QKD was studied previously for modulated coherent-state \cite{Lev10,Jou12} and entanglement-based \cite{Fur12,Ebe13} protocols. The channel estimation in the current state-of-the-art implementation \cite{Jou13} is based on revealing half of the pulses publicly and estimating from that the channel noise and transmittance. However, the channel estimation strategy was not optimized for the given quantum-state resource and other parameters of the scheme, and information on the effect of channel estimation on the squeezed-state protocol appears to be lacking, which limits the performance of the standard protocol. 

In the current paper we suggest a novel channel estimation strategy based on double Gaussian modulation of the coherent or squeezed states of light. We show that this method can bypass the trade-off  between estimation and information transfer and that the optimized protocol can approach the effectiveness of the ideal case. We also study the protocol based on nonclassical squeezed states and show the advantage of using such states in CV QKD when channel estimation and finite-size effects are considered. In summary, we present a CV QKD protocol that is practically applicable in long-distance channels thanks to the optimal use of the resources.

The paper is organized as follows. First, we describe in Sec. II the basic concepts of CV QKD: the model used, the standard method for channel estimation, and the calculation of the achievable key rate. In Sec. III, we optimize the standard method of channel estimation used for CV-QKD incorporating finite-size effects. Then in Sec. IV, we calculate the theoretical limits of CV QKD. This leads us to the concept of a double-modulation protocol (Sec. V) which can approach the optimal performance for long distances. Finally, in Sec. VI, we draw conclusions.
 


\section{Preliminaries}

\subsection{The model used} 

\begin{figure}[!ht]
\begin{center}
  \includegraphics[width=8.5cm]{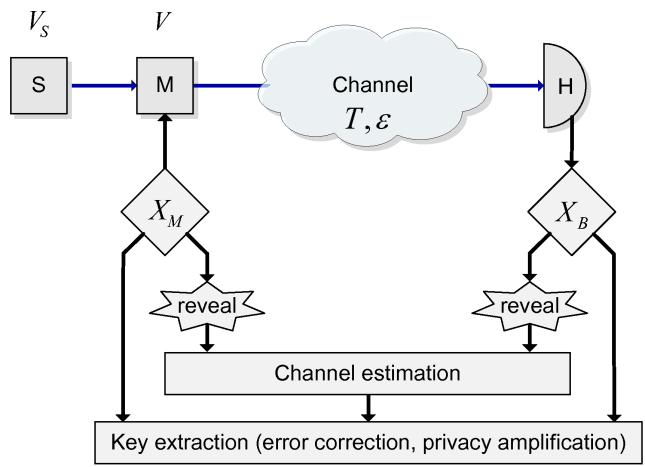}
 \caption{Prepare-and-measure CV-QKD with single modulation: using a Gaussian source (S) and modulator (M) at Alice's side and a homodyne detector (H) at Bob's side. One subset of the data is used for estimation; the other subset is used for key distribution. 
\label{method}}
\end{center}
\end{figure}

We consider the generic CV QKD protocol, when Alice sends CV quantum states of light to Bob through a channel, that is under the control of a potential eavesdropper Eve. The resource states are Gaussian squeezed or coherent states; Alice applies a Gaussian displacement operation to them, while Bob performs homodyne detection on the other end of the channel using also an intense reference pulse sent by Alice (Fig. \ref{method}). After the transmission is done, they reveal a subset (chosen randomly) of their measurement data, estimate the channel parameters from that, and use that information to upper-bound the information Eve could get. Finally they use the other, unrevealed subset of data to extract the key, performing also error correction and privacy amplification. 

In this case the transfer of quadrature variables through a lossy and noisy channel can be described in the Heisenberg picture by the following evolution:
\begin{equation}\label{def_channel}
x_B=\sqrt{T}\cdot (x_S+x_M)+\sqrt{1-T} \cdot x_0+x_\eps, 
\end{equation}
where all variables are normally distributed with zero mean, except for the fixed transmittance parameter $T \in [0,1]$. The operator $x_B$ represents the measured quadrature at Bob's side. The operator $x_A=x_S+x_M$ represents the quadrature of a state Alice sent into the channel, where $x_S$ comes from the quantum fluctuation of the resource state [with variance $\Var(x_S)=V_S$], while $x_M$ comes from the modulation of the state [$\Var(x_M)=V$]. Note that $V_S=1$ (shot noise unit) for coherent states and $V_S<1$ for squeezed states. $x_0$ corresponds to the vacuum state [$\Var(x_0)=1$], while $x_{\eps}$ is a Gaussian excess noise with variance $\Var(x_{\eps})=V_{\eps}$.  
In practice the values of $x_S$, $x_0$, and $x_\eps$ are all unknown to Alice and Bob, so they can treat all of them as a noise. So we can rewrite (\ref{def_channel}) in a much simpler form: 
\begin{equation}\label{def_channel_simple}
x_B=\sqrt T \cdot x_M+x_N,
\end{equation}
where $x_N$ is the aggregated noise with zero mean and variance $V_N=1+V_\eps+T(V_S-1)$.



\subsection{The standard method for channel estimation}
We suppose that the variance of the modulation ($V$) is known, that is, the channel can be parametrized using two unknown parameters: $T$ and $V_{\eps}$. 

Let us suppose that the estimation is made using $m$ Gaussian states. Denote the realizations of $x_M$ and $x_B$ with $M_i$ and $B_i$ ($i\in \{1,2,\dots,m\}$) respectively. 
We know that the covariance of $x_M$ and $x_B$ is $\Cov(x_M,x_B)=\sqrt{T}\cdot V=:C_{MB}$. We can estimate the value of $T$ from
\begin{equation}\label{est_eta}
\hat T=\frac{1}{V^2}\cdot \left(\widehat{C_{MB}}\right)^2,
\end{equation}
where we use the maximum likelihood estimator
\begin{equation}\label{est_Cov}
 \widehat{C_{MB}}=\frac{1}{m}\sum_{i=1}^m M_i B_i.
\end{equation}

On the other hand, the estimator of $V_\eps$ can be expressed using the maximum likelihood estimator of $V_N$ substituting the real value of $T$ with its estimator from (\ref{est_eta}):
\begin{equation}\label{hatVeps}
\hat V_\eps=\frac{1}{m} \sum_{i=1}^m (B_i-\sqrt{\hat T} M_i)^2+\hat T(1-V_S)-1
\end{equation}

Let us notice that to estimate $V_\eps$ either Alice or Bob should reveal the measurement data, so these states can not be used for key distribution.


\subsection{The achievable secret key rate}
We will use reverse reconciliation to obtain a secret key for large distances, i.e., the common key is based on Bob's state, which Alice and Eve try to guess. Then in the asymptotical case the key rate is \cite{Lod10}
$$
K_{\infty}(T,V_\eps)=\beta I(A:B)-S(B:E),
$$
where $\beta \in [0,1]$ is the reconciliation efficiency, $I(A:B)$ is the mutual information of Alice and Bob, while $S(B:E)$ is the maximal information Eve can retain about Bob's state.

The true values of $T$ and $V_\eps$ are unknown, so for implementing a secure CV QKD protocol we need to set confidence intervals for both $T$ and $V_\eps$ with a low probability of error $\delta$(see Appendix C). In order not to underestimate the eavesdropping, the key rate must be minimized considering every possible combination of $T$ and $V_\eps$ from the given confidence intervals.
Numerical calculations (and intuition) suggest that in the most pessimistic case, we should use the lower bound ($T^{low}$) of the confidence interval for $T$ and the upper bound ($V_\eps^{up}$) for $V_\eps$. 

We can obtain the following key rate incorporating finite-size effects \cite{Lev10}:
\begin{equation}\label{key_rate}
K=\frac{n}{N}\cdot \bigg[K_{\infty}(T^{low},V_\eps^{up})-\Delta(n)\bigg],
\end{equation}
where $n$ is the number of Gaussian states used for secret key transmission and $\Delta(n)$ is a correction term for the achievable mutual information in the finite case \cite{Sca08}. Note that if we use $m=r \cdot N$ states for estimation we will have $n=(1-r) \cdot N$ states for key distribution.  

The mutual information reads
$$
I(A:B)=\frac12 \log_2\Bigg(1+\frac{V\cdot T}{V_N}\Bigg).
$$


We suppose that Eve performs a collective Gaussian attack on the signal pulse (the reference pulse is usually much stronger than the signal pulse). In this case, the upper bound of the information which is available to Eve on Bob's measurement results is given by the Holevo information, that is, the difference between two von Neumann entropies:
$$
S(B:E)=S_E-S_{E|B},
$$
where $S_E$ is the von Neumann entropy of the eavesdropper's state, while $S_{E|B}$ denotes the  von Neumann entropy of the eavesdropper's state conditioned on Bob's measurement.

\begin{figure}[!ht]
\begin{center}
    \includegraphics[width=8.5cm]{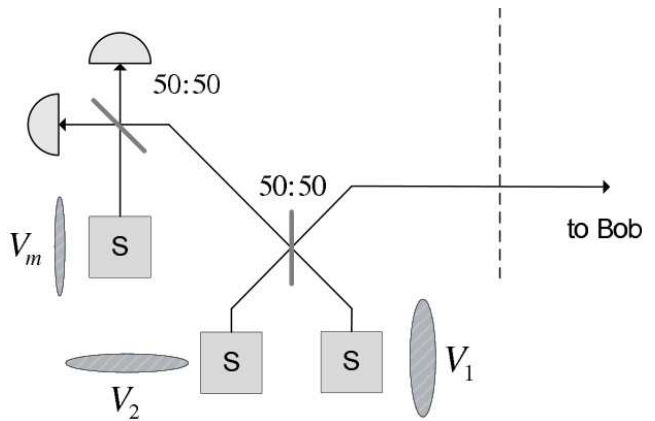}
 \caption{The equivalent entanglement-based scheme that is used for calculating the Holevo information instead of the prepare-and-measure scheme \cite{Use11}. \label{fig_entangled}}
\end{center}
\end{figure}

In the general case the channel noise is assumed to be under full control of Eve, so in order to calculate these entropies we use an equivalent entanglement-based scheme (Fig. \ref{fig_entangled}) and purification method \cite{Use11}. It is equivalent in the sense that the states and conditional states (conditioned on Alice's measurements) sent to Bob through the channel have the same distribution as in the prepare-and-measure scheme. A generalized entanglement-based scheme is used, which corresponds to the preparation of arbitrarily squeezed states and arbitrary modulation applied to them. This scheme differs from the standard entanglement-based schemes by the presence of an additional squeezed state coupled to a signal prior to measurement on the sending side, and also by the unbalanced preparation of the entangled state.

With this equivalence we can substitute a prepare-and-measure scheme with arbitrary squeezing of the signal states and modulation variances with an entanglement-based scheme. The calculation of the Holevo information in the entanglement-based scheme is straightforward from here using purification. From the fact that Eve is able to purify the state shared between Alice and Bob it follows that $S_E=S_{AB}$ and $S_{E|B}=S_{A|B}$. Both entropies can be calculated from the symplectic eigenvalues of the covariance matrices of two-mode Gaussian states. That is, we can express them as an (enormously long) analytic formula of parameters and we can use it efficiently for the numerical optimization of parameters.



Finally, for the sake of simplicity we will use the approximate formula for $\Delta(n)$ obtained in \cite{Lev10}:
$$
\Delta(n)\approx 7 \sqrt{\frac{\log_2(2/\delta^*)}{n}},
$$
where $\delta^*$ is the probability of error during privacy amplification.



\section{Optimization of the standard method}
The key rate (\ref{key_rate}) can be calculated for fixed parameters: it depends only on the values of $T$, $V_\eps$, $N$, $\beta$, $r$, $V$, and $V_S$. Let us investigate thoroughly how the key rate depends on these parameters.

The first two are given; they are the parameters of the channel. One can estimate them with the estimators in Eqs. (\ref{est_eta}) and (\ref{hatVeps}). We can approximate their variances with (see Appendix A)

\begin{equation}\label{var_hatt}
  \Var(\hat T)\approx \frac{4}{m} \cdot T^2 \left(2+\frac{V_N}{T V}\right)=:\sigma_1^2
\end{equation}
and 
\begin{equation}\label{var_veps}
  \Var(\hat V_\eps)\approx \frac{2}{m}\cdot V_N^2+(1-V_S)^2\cdot \sigma_1^2=:s_1^2.
\end{equation}
From that we can calculate the expected values of $T^{low}$ and $V_\eps^{up}$ in Eq. (\ref{key_rate}), so we can calculate the key rate for any set of parameters in advance numerically.
Otherwise, for an optical fiber of length $d$ (in km) we will use $T=10^{-0.2 d/10}$. In the literature it is usually assumed that the excess noise in an optical fiber is proportional to the transmittance \cite{Jou12,Lev10,Fur12}. So we used this assumption in our numerical calculation as well, having $V_{\eps}=T \cdot \eps$, with $\eps=0.01$.

$N$ is the size of the blocks; it is in general predetermined, but for practical applications it is reasonable to assume short blocks. In the current state-of-the-art realization \cite{Jou13} $N=10^8$ and $N=10^9$ were used. 

$\beta$ is the efficiency of the information reconciliation, which depends on the performance of the algorithms being used and on the achieved signal-to-noise ratio. Recently efficient postprocessing algorithms were developed \cite{Jou11}; thus we will use a realistic $\beta=0.95$.

\begin{figure}[!t]
\begin{center}
    \includegraphics[width=8.5cm]{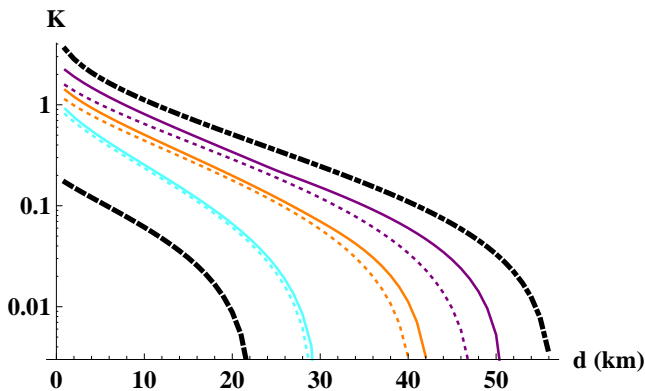}
  \caption{(Color online) The secure key rate of different methods: for coherent states [cyan (light gray), $V_S=1$], moderate squeezing [orange (medium gray), $V_S=0.5$], and strong squeezing [purple (dark gray), $V_S=0.1$], for optimized single-modulation (dotted lines) and double-modulation (solid lines) schemes as a function of distance $d$ for $\eps=0.01$, $\beta=0.95$, and $N=10^6$. For comparison we plotted also the key rate yielded by the  current state-of-the-art technique (thick black dashed line) and the best theoretically possible upper limit $K^{th}$ (thick black dash-dotted line). \label{dist}}
\end{center}
\end{figure}

\begin{figure}[!t]
\begin{center}
    \includegraphics[width=8.5cm]{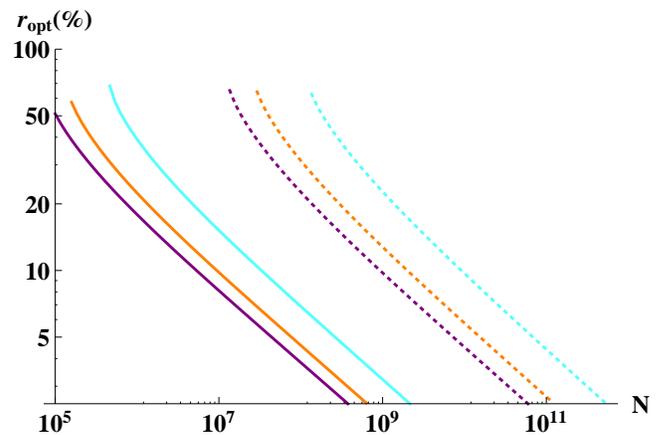}
 \caption{(Color online) The optimal ratio used for estimation for the single modulation method: for coherent states [cyan (light gray), $V_S=1$], moderate squeezing [orange (medium gray), $V_S=0.5$], and strong squeezing [purple (dark gray), $V_S=0.1$], for a long distance $T=0.03$ ($d\approx$ 76 km, dotted lines) and a short distance $T=0.3$ ($d\approx$ 26 km, solid lines) as a function of block size $N$ with $\eps=0.01$ and $\beta=0.95$. \label{fig_r}}
\end{center}
\end{figure}

From (\ref{key_rate}) we can see that the key rate will be higher if more states are used for key distribution ($n$). But at the same time that means fewer states for estimation ($m$) and results in an inaccurate estimation of channel parameters. In the case of perfect reconciliation $V$ should be as large as possible, but in a realistic case $V$ must be limited. In fact, there exist optimal values for $r$ and $V$ which maximize the key rate. But so far the variances of the parameter estimators have not been obtained in general, only for given measurement outcomes. This means that designing an experiment by optimizing the available parameters was impossible; some fixed parameters were used instead, e.g., $V=1.5$ \cite{Jou12} and $r=1/2$ \cite{Lev10,Jou13}. However, since we know the variances of the estimators in advance, the optimal setting can be calculated by numerical optimization and a significant improvement is achieved in the key rate (Fig. \ref{dist}, thick black dashed vs purple dotted line). The optimal $V$ will be close to the one obtained in the asymptotical case (without any finite-size effects) \cite{Use11}. While from Fig. \ref{fig_r} we can see that the optimal $r$ will be close to $50\%$ only if the key rate is close to zero. If that is not the case, the optimal ratio will be below $50\%$ and show a linear correlation with the block size ($N$) on a log-log plot.  That is, the optimal ratio $r_{opt}$ will have the form of $r_{opt}\approx \alpha x^\gamma$ with the actual value of $\alpha$ and $\gamma$ depending on the parameters. In general $\alpha$ will be lower for smaller distances and higher levels of squeezing, while $\gamma\approx -0.35$ (the lines on Fig. \ref{fig_r} are nearly parallel).

Finally, $V_S$ is a squeezing parameter of the source, which can be set in state preparation. If we compare the performance using coherent states ($V_S=1$), moderately squeezed states ($V_S=0.5$, i.e., 3 dB squeezing) and strongly squeezed states ($V_S=0.1$, i.e., 10 dB squeezing), we can see a substantial improvement due to squeezing  (Fig. \ref{dist}, dotted lines), even for the moderately squeezed states. With the use of squeezing the protocol can achieve reasonable distances for even  values of $N$ as low as $10^6$.



\section{Limitations on the key rate}
One of the main limiting factors compared to the asymptotical case comes from the fact that there will be a security break if the asymptotical key rate drops below $\Delta(n)$. This quantity is of order $c/\sqrt{n} ~ (\textrm{with } c\in \mathbb R_+)$ which results in a substantial restriction on achievable distance even for large values of $n$. To get a higher key rate one can try to improve the coefficient $c$ using theoretical considerations \cite{Fur12}. For a given function $\Delta$ the possible improvement comes from using as large $n$ as possible, i.e., $n=N$.

\begin{figure}[!t]
\begin{center}
    \includegraphics[width=8.5cm]{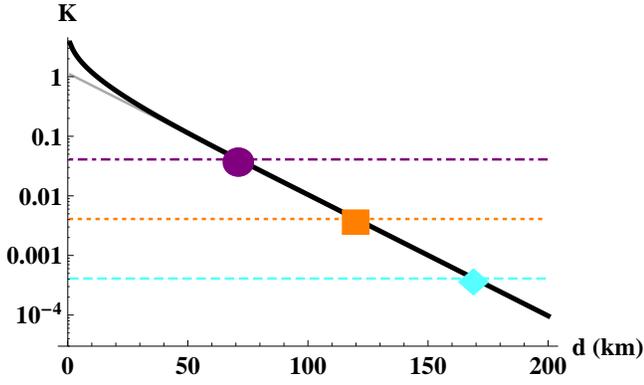}
 \caption{(Color online) The asymptotical key rate $K_\infty (T,V_\eps)$ (thick black solid line) using optimal modulation and infinitely strong squeezing with $\eps=0.01$ and $\beta=0.95$, its exponential approximation (solid gray line), and the level of $\Delta(N)$ for $N=10^6$ (dash-dotted), $N=10^8$ (dotted), and $N=10^{10}$ (dashed). If the asymptotical key rate drops below $\Delta(N)$ (large circles), it is impossible to obtain a positive key rate. \label{fig_max_dist}} 
\end{center}
\end{figure}

Actually, from the $K_{\infty}\ge c/\sqrt{N}$ restriction one can easily obtain that the maximally achievable distance is in the best case a linear function of $\log_{10} N$: if we take ten times larger block sizes, we can expect an improvement of about 25 km (see Fig. \ref{fig_max_dist} and Appendix D).

In practical situations the transmittance $T$ can be estimated quite promptly, but that is not true for the excess noise $V_\eps$. For large distances $V_\eps=T\cdot \eps$ will be very small; nevertheless $V_\eps^{up}$ will be large since the estimator $\hat V_\eps$ will have a large standard deviation. Simple calculations show that we have
\begin{equation}\label{Ve_up}
V_\eps^{up} \approx \sqrt{2} \cdot \frac{1+V_\eps+T (V_M+V_S-1)}{\sqrt{m}},
\end{equation}
where $V_M$ is the conditional modulation of Alice: $V_M=0$ if Alice reveals the modulation data for Bob; $V_M=V$ if Alice does not reveal the modulations. The theoretical lower bound is 
\begin{equation}\label{Ve_up_th}
V_\eps^{up} \ge V_\eps^{th} = \sqrt{2} \cdot \frac{1+V_\eps-T }{\sqrt{N}},
\end{equation}
and is fulfilled if Alice shares all the modulation for channel estimation and uses infinitely squeezed states.

From these two observations we can get the theoretical maximum for the key rate (\ref{key_rate}) in the finite case:
\begin{equation}\label{key_rate_th}
K^{th}=K_{\infty}(T,V_\eps^{th})-\Delta(N).
\end{equation}

Unfortunately, this is impossible to achieve since all states would have to be used for both optimal key distribution and optimal channel estimation at the same time (the latter means revealing the modulation for all states).


\section{Double-modulation method}\label{double}

Using the standard method we set $V_M=0$ in (\ref{Ve_up}), that is, Alice reveals the exact values of modulation. However, she uses only half of the states, which still results in a much higher value of $V_\eps^{up}> \sqrt{2} \cdot V_\eps^{th}$. But for large distances $T$ becomes very small, so the term in parentheses in (\ref{Ve_up}) will not have a large effect. This observation has caused us to take a different approach to the problem: Alice should not share the modulation data at all. In this case Alice can use all the states for estimation, which for large distances results in a value much closer to the theoretical limit: $V_\eps^{up} \rightarrow V_\eps^{th}$, if $T\rightarrow 0$. Let us note that in this case, besides having a better estimate of $V_\eps$, one can use twice as many states for key distribution, too.

\begin{figure}[!t]
\begin{center}
  \includegraphics[width=8.5cm]{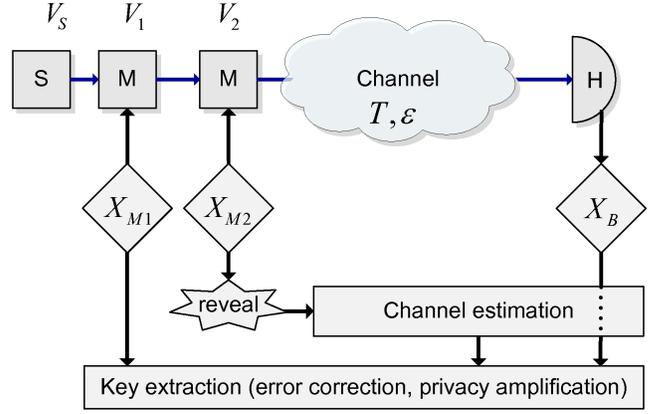}
 \caption{Prepare-and-measure CV-QKD with double modulation: using a Gaussian source (S) and two modulators (M) at Alice's side and a homodyne detector (H) at Bob's side. 
One modulation can be used for estimation, and the other modulation for key distribution; hence all the states can be used for both estimation and key distribution at the same time. 
\label{method2}}
\end{center}
\end{figure}

To actually achieve this effect, we also need a method to estimate $T$ without revealing the modulation data. That led us to using two consecutive modulations on Alice's side (see Fig. \ref{method2}). After finishing the transmission Alice reveals for each state the second modulation ($x_{M2}$). Then Bob using these public data and his (secret) measurement data can estimate $T$ and $V_\eps$ (attributing the first modulation of Alice to the noise of the source).  In this way the first modulation ($V_1$) and Bob's measurement remain secret, so they can be used for key distribution as in the standard case (having an additional noise coming from the second modulation, but since that is revealed publicly it can be simply eliminated from the process).


\begin{figure}[!t]
\begin{center}
 \includegraphics[width=8.5cm]{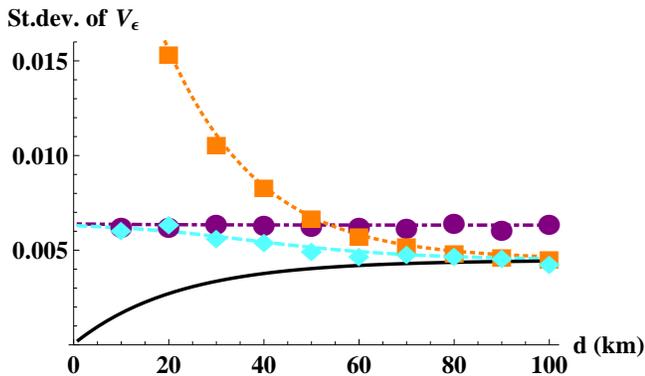}
\caption{(Color online) The approximated standard deviation (lines) and the real values calculated from numerical simulation (symbols). The values of $s_1$ [standard method, purple (dark gray) dash-dotted line), $s_2$ [double modulation, orange (medium gray) dotted line), and $s_3$ [modified double modulation, cyan (light gray) dashed line] for $N=10^5$, $r=0.5$, $V=V_1=3$, $V_2=10$, and $\eps=0.01$ using coherent states ($V_S=1$) show a good match with the empirical standard deviation of different estimators of $V_\eps$ (circles, squares, and diamonds, respectively) averaged from $10^3$ different realizations. The theoretical lower limit from $V_\eps^{th}$ (black solid line) is also plotted for comparison. \label{fig_stdev}}
\end{center}
\end{figure}


Then the evolution can be written in the form of
\begin{equation}\label{def_channel2}
x_B=\sqrt{T}\cdot (x_S+x_{M1}+x_{M2})+\sqrt{1-T} \cdot x_0+x_\eps.
\end{equation}

Since Alice reveals only the values of the second modulations ($x_{M2}$), the first modulation acts as a noise in the estimation process; thus we can rewrite (\ref{def_channel2}) in the following form:
\begin{equation}\label{def_ch2}
x_B=\sqrt T \cdot x_{M2}+x_N^*,
\end{equation} 
where $x_N^*$ is the aggregated noise with variance $V_N^*=1+V_\eps+T(V_1+V_S-1)$.

That means that it is the same evolution as for a single modulation [see Eq. (\ref{def_channel_simple})], we need only to change the variance $V$ to $V_2$, $V_N$ to $V_N^*$, and $m$ to $N$. So by substitution in Eqs. (\ref{var_hatt}) and (\ref{var_veps}) we can easily obtain the approximated variance for the estimator of $T$:

\begin{equation}\label{vareta2}
  \sigma_2^2:=\frac{4}{N} \cdot T^2 \left(2+\frac{V_N^*}{T V_2}\right),
 \end{equation}

and the approximated variance for the estimator of $V_\eps$:

 \begin{equation}\label{varVeps2}
  s_2^2:=\frac{2}{N}\cdot (V_N^*)^2+(V_1+V_S-1)^2\cdot \sigma_2^2.
 \end{equation}

This formula shows the effects described above. The aggregated noise $V_N^*$ and the factor $(V_1+V_S-1)^2$ will be larger here than in the standard case. That results in a worse key rate for small distances. But if $T \rightarrow 0$, then $V_N^* \rightarrow V_N$ and $\sigma_2^2 \rightarrow 0$, so these negative effects will disappear, thus from using more states for estimation, one can get an even better estimation (see Fig. \ref{fig_stdev}, purple dash-dotted and orange
dotted lines). Moreover, for large distances the variance of the estimator gets close to the theoretical optimum (see Fig. \ref{fig_stdev}, orange dotted and black solid lines). Note that $V_2$ plays a role only in the estimation of $T$. Larger values of $V_2$ always produce lower values of $\sigma_2$ (so also lower values of $s_2$), but the improvement saturates quickly.

\begin{figure}[!t]
\begin{center}
  \includegraphics[width=8.5cm]{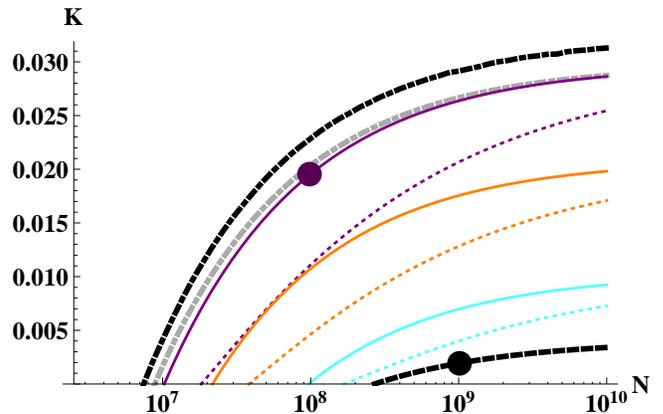}
 \caption{(Color online) The secure key rate of different methods: for coherent states [cyan (light gray), $V_S=1$], moderate squeezing [orange (medium gray), $V_S=0.5$], and strong squeezing [purple (dark gray), $V_S=0.1$], for the optimized single-modulation (dotted lines) and double-modulation (solid lines) schemes as a function of block size $N$ in the case of $\eps=0.01$, $\beta=0.95$, and $T=0.03$ ($d=76$ km). For comparison we plotted also the key rate using the current state-of-the-art technique (thick black dashed line), and the best theoretically possible upper limit $K^{th}$ using $V_S=0.1$ (thick gray dash-dotted line), and infinitely strong squeezing (thick black dash-dotted line). \label{N_b}}
\end{center}
\end{figure}


That is, we obtained a method which estimates the excess noise efficiently. But in the meantime, besides using all the states for estimation, one can use all states for key distribution purposes too. That will result in a key rate (Fig. \ref{N_b}) approaching the theoretical limit described in (\ref{key_rate_th}). It is important to mention that even with a feasible level of squeezing we can obtain a key rate close to the theoretical optimum corresponding to infinitely strong squeezing.

If $T$ is not close to zero, the above method will not be efficient. But in that case Alice can share some data from the first modulation too. That is, in this case we will have the same situation as in the standard method; there is only an additional layer of noise for every state, which will be revealed and used for better estimation (see Appendix B). 

We can optimize the ratio of shared modulation ($r$) and get a slightly improved key rate compared to the single-modulation scheme (Fig. \ref{dist} dotted vs solid lines). Not surprisingly the improvement is higher for larger distances. The optimal ratio $r$ becomes zero between $T=0.1$ and $T=0.3$ (that is between 25 and 50 km) depending on the parameters. Thus for large distances, to get the optimal performance one should indeed use each state for estimation and key distribution too. 

The implementation of this protocol is simple. From some rough estimation of channel parameters (e.g., from earlier results) Alice chooses a modulation variance that is close to optimal. After Bob received the states Alice reveals all her data of the second modulations. From this they will have a proper estimation of the channel parameters and they can calculate the optimal ratio $r$ numerically. If it is positive, Alice chooses $r\cdot N$ states randomly and reveals their first modulation, which will be incorporated to obtain a more accurate estimation. Then they continue the protocol as in the standard case.


\section{Conclusion and discussion}
We proposed a feasible double modulation quantum key distribution scheme which uses each state for both channel estimation and key distribution purposes. We presented a simple theoretical maximum for the key rate and the achievable distance for CV-QKD protocols. We showed that our method can greatly improve the key rate and the maximal distance, moreover, it can approach the theoretical limit for long distances.

The improvement comes from three factors: optimizing the parameters, using squeezed states, and using the double-modulation method (see Fig. \ref{N_b} for the different effects). The full optimization for the finite-size case was feasible because we obtained formulas as a simple function of the model parameters for a good approximation of the standard deviation of the channel parameter estimators. Using squeezed states instead of coherent states makes the largest improvement, which also theoretically clarifies that the result obtained in \cite{Use11, Mad12} remains valid if we take finite-size effects into account (this surviving effect is far from trivial in general). Finally, there is the double modulation method, which allows us to approach the theoretical limit for large distances, proving that the above ideas are enough to realize a nearly optimal CV QKD scheme. For comparison, we obtained a key rate about ten times higher than the current state-of-the-art implementation \cite{Jou13} even if we use ten times shorter block sizes (Fig. \ref{N_b}, large circles).

Note that the results were obtained for a specific protocol; however, the same concepts can be applied to different and more complex settings of CV QKD, which implies that they might produce a significant improvement in practical applications of quantum communication.

\subsection*{Acknowledgments}
V.C.U. and L.R. acknowledge the Project No. 13-27533J of GA\v CR. The research leading to these results has received funding from the EU FP7 under Grant Agreement No. 308803 (Project BRISQ2), co-financed by M\v SMT \v CR (7E13032).




%

\subsection*{Appendix A: Variances for the single-modulation method}

To obtain the variances of the estimators of interest we first calculate the variance of $\widehat{C_{MB}}=\frac{1}{m}\sum_{i=1}^m M_i B_i$.

This will be an unbiased estimator of $C_{MB}$, since 
$$
\bE(\widehat{C_{MB}})=\frac{1}{m}\sum_{i=1}^m \bE(M_i B_i)=\bE(x_M \cdot x_B)=
$$
$$
=\bE\Bigg(x_M\cdot \bigg(\sqrt{T}\cdot x_M+x_N\bigg)\Bigg)=\sqrt{T} ~\bE( x_M^2)=C_{MB}.
$$

On the other hand, we have 
$$
\Var(\widehat{C_{MB}})=\frac{1}{m^2}\sum_{i=1}^m \Var(M_i B_i)=\frac{1}{m} ~\Var(x_M\cdot  x_B)=
$$
$$
=\frac{1}{m} \Var\bigg(x_M\cdot (\sqrt{T}\cdot x_M+x_N)\bigg)=
$$
$$
=\frac{1}{m} \bigg(T~ \Var(x_M^2)+\Var(x_M x_N)\bigg)=\frac{1}{m} (T \cdot 2V^2+V \cdot V_N),
$$
where we used the definition of $x_B$, the moments of a normal distribution and the fact that for independent random variables with zero mean the variance can be factorized. 

From this we can finally obtain the variance
\begin{equation}
V_{\mathrm{Cov}}:=\Var(\widehat{C_{MB}})=\frac{1}{m}\cdot T V^2 \bigg(2+\frac{V_N}{T V}\bigg),
\end{equation}
which is of order $1/m$, so it is a consistent estimator (i.e., $\widehat{C_{MB}}\rightarrow C_{MB}$ if $m$ goes to infinity).


Now we can obtain the properties of the estimators used in the standard method. 
First we calculate the properties of the estimator of the transmittance: $\hat T=\frac{1}{V^2}\cdot \left(\widehat{C_{MB}}\right)^2$.

If we rearrange the second term:
$$
(\widehat{C_{MB}})^2=V_{\mathrm{Cov}}\cdot \left(\frac{\widehat{C_{MB}}}{\sqrt{V_{\mathrm{Cov}}}}\right)^2,
$$
then the last expression will be noncentrally $\chi^2$ distributed:
$$
\left(\frac{\widehat{C_{MB}}}{\sqrt{V_{\mathrm{Cov}}}}\right)^2 \sim \chi^2\left(1,\frac{C_{MB}^2}{V_{\mathrm{Cov}}}\right).
$$
From that we can obtain the mean
$$
\bE(\hat T)=\frac{V_{\mathrm{Cov}}}{V^2}\cdot \bE\left(\frac{\widehat{C_{MB}}}{\sqrt{V_{\mathrm{Cov}}}}\right)^2=\frac{V_{\mathrm{Cov}}}{V^2}\cdot \left(1+\frac{C_{MB}^2}{V_{\mathrm{Cov}}}\right)=
$$
$$
=\frac{V_{\mathrm{Cov}}+C_{MB}^2}{V^2}=\frac{V_{\mathrm{Cov}}+T V^2}{V^2}=T+\frac{V_{\mathrm{Cov}}}{V^2}=T+O(1/m),
$$
and the variance
$$
\Var(\hat T)=\frac{V_{\mathrm{Cov}}^2}{V^4}\cdot\Var\left(\frac{\widehat{C_{MB}}}{\sqrt{V_{\mathrm{Cov}}}}\right)^2=\frac{V_{\mathrm{Cov}}^2}{V^4}\cdot2\left(1+2\frac{C_{MB}^2}{V_{\mathrm{Cov}}}\right)=
$$
$$
=\frac{2V_{\mathrm{Cov}}\cdot(V_{\mathrm{Cov}}+2C_{MB}^2)}{V^4}=\frac{2V_{\mathrm{Cov}}\cdot 2C_{MB}^2}{V^4}+O(1/m^2),
$$
where we have used that $V_{\mathrm{Cov}}$ is of order $1/m$. We can rewrite the variance of $\hat T$ in the following form:
 \begin{equation}\label{vareta}
  \sigma_1^2:=\Var(\hat T)=\frac{4}{m} \cdot T^2 \left(2+\frac{V_N}{T V}\right)+O(1/m^2).
 \end{equation} 

We can see that $\hat T$ is not unbiased, only asymptotically unbiased. But the bias is of order $1/m$, while its standard deviation $\sigma_1$ is of order $1/\sqrt{m}$, meaning that the magnitude of the bias is negligible compared to it, so $\hat T$ can be used to estimate $T$. Note that in further calculations it suffices to use only the first-order approximation, since typically $m>10^5$, so the second term will be negligible.


Now we can calculate the properties of the estimator of the excess noise: $\hat V_\eps=\frac{1}{m} \sum_{i=1}^m (B_i-\sqrt{\hat T} M_i)^2+\hat T(1-V_S)-1$. We can calculate its variance by substituting the definition of the estimator $\sqrt{\hat T}$ into the sum, but at the end there would not be much difference from using simply $\sqrt{\hat T}=\sqrt{T}$, i.e., assuming that $\hat T$ has negligible variance. The reason behind that is that for large values of $m$, the estimator $\sqrt{\hat T}$ will be very close to its real value; thus the main source of uncertainty in the sum comes from the random variables $B_i$ and $M_i$. So for the sake of simplicity, we will present in the following the simpler analysis.

It is easy to see that $B_i-\sqrt{T} M_i$ is normally distributed with variance $V_N$. So $Y:=\sum_{i=1}^m \left(\frac{B_i-\sqrt{T} M_i}{\sqrt{V_N}}\right)^2$ will be $\chi^2$ distributed: $Y \sim \chi^2(m)$, with $\bE(Y)=m$ and $\Var(Y)=2 m$. 

Then $\sum_{i=1}^m (B_i-\sqrt{\hat T} M_i)^2$ can be approximated by $V_N\cdot Y$ for large values of $m$ and we can obtain
$$
\bE (\hat V_\eps)=\bE \left(-1+\hat T(1-V_S)+\frac{1}{m} \sum_{i=1}^m (B_i-\sqrt{\hat T} M_i)^2\right)\approx
$$
$$
\approx -1 + T(1-V_S)+\frac{1}{m}V_N \cdot \bE(Y)=V_\eps.
$$
To calculate the variance we assume that $\hat T$ and $(B_i-\sqrt{\hat T} M_i)^2$ are independent. This is not exactly true, but numerical simulations show us that this assumption still gives us a good approximation in the current situation (see Fig. \ref{fig_stdev}). Hence we calculate the variances independently for each term:

$$
\Var \left(-1+\hat T(1-V_S)+\frac{1}{m} \sum_{i=1}^m (B_i-\sqrt{\hat T} M_i)^2\right)\approx
$$
$$
\approx(1-V_S)^2 \Var(\hat T)+\frac{1}{m^2} V_N^2 \Var(Y).
$$

That is, in other words, we can approximate the variance of $\hat V_\eps$ with
 \begin{equation}\label{varVeps}
  \Var(\hat V_\eps)\approx \frac{2}{m}\cdot V_N^2+(1-V_S)^2\cdot \sigma_1^2=:s_1^2.
 \end{equation}

Let us note that $\sigma_i^2$ is of order $1/m$, so both terms in (\ref{varVeps}) will be of order $\frac{1}{m}$, too. For coherent states $s_1^2$ will be constant, $\frac{2(1+V_\eps)^2}{m}$ (see Fig. \ref{fig_stdev}, purple dash-dotted line).


\subsection*{Appendix B: Variances for the modified double-modulation method}

To get an accurate estimation for low distances in double-modulation settings, Alice needs to share some of the first modulations too. Let us suppose that Alice reveals $m=r\cdot N$ first modulations. Then one can calculate an estimation of $T$ and $V_\eps$ from both subsets of the states independently. 

For the $(1-r)\cdot N$ states for which Alice reveals only the second modulation we can use the same calculation as described in Sec. \ref{double} around Eq. (\ref{varVeps2}) [with the small difference that we should use $(1-r)\cdot N$ instead of $N$]. While for the $r\cdot N$ states for which Alice reveals both modulations, the evolution is
\begin{equation}\label{def_ch3}
x_B=\sqrt T \cdot (x_{M1}+x_{M2})+x_N.
\end{equation} 
That is, it is the same situation as in the standard method, with the difference that there should be $V_1+V_2$ instead of $V$ in the appropriate formula.

One can easily verify that if there are two different estimators $\hat x_1$ and $\hat x_2$ with variances $W_1$ and $W_2$, then the best linear estimator $\alpha \cdot \hat x_1 + (1-\alpha) \cdot \hat x_2$ is yielded by setting $\alpha=\frac{W_2}{W_1+W_2}$. In this case the minimal variance is
$$
\mathrm{opt}(W_1,W_2):=\frac{W_1\cdot W_2}{W_1+W_2}=\frac{1}{\frac{1}{W_1}+\frac{1}{W_2}}.
$$

Using this result we can construct the best linear estimator from the two independent estimators (corresponding to two subsets of states) and we can obtain the variance of the estimator of $T$:

\begin{widetext}
\begin{equation}\label{vareta3}
  \sigma_3^2:=\mathrm{opt}\bigg(\frac{4}{(1-r)N} \cdot T^2 \left(2+\frac{V_N^*}{T V_2}\right)~,~\frac{4}{r N} \cdot T^2 \left(2+\frac{V_N}{T (V_1+V_2)}\right)\bigg)
 \end{equation}

and the variance of the estimator of $V_\eps$:

 \begin{equation}\label{varVeps3}
  s_3^2:=\mathrm{opt}\bigg(\frac{2}{(1-r)N}\cdot (V_N^*)^2+(1-V_S-V_1)^2\cdot \sigma_3^2 ~,~ \frac{2}{r N}\cdot V_N^2+(1-V_S)^2\cdot \sigma_3^2 \bigg).
 \end{equation}
\end{widetext}

This method combines the advantages of the single- and double-modulation methods (see Fig. \ref{fig_stdev}, cyan dashed line). It provides the optimal estimation, converges to the standard method for low distances, while for large distances, converges to the double-modulation method.

It is important to notice that in our calculations we have used approximated variances, but Fig. \ref{fig_stdev} shows us that these approximations are close to the empirical variances even for $N=10^5$. Therefore we can use these approximate formulas to numerically calculate the key rate.

We should also note that in the case of squeezed sources a moderate improvement in standard deviations can be experienced, but that does not change fundamentally the relations discussed above.


\subsection*{Appendix C: Confidence intervals}

Let us suppose that $X$ is an estimator of interest and it is normally distributed with mean $\mu$ and standard deviation $\sigma$. We are looking for a symmetric confidence interval (around $\mu$), and denote the significance level of the confidence interval with  $\delta$, that is,
\begin{equation}\label{error}
P\left( \mu-\alpha < X < \mu + \alpha \right )=1-\delta.
\end{equation}

Therefore the probability of an estimate above the upper bound is $\delta/2$:
$$
\delta/2 = P\left( X > \mu + \alpha \right )=P\left( \frac{ X - \mu}{\sigma}> \frac{\alpha}{\sigma} \right )=
$$
$$
=P\left(Y>\frac{\alpha}{\sigma}\right)=1-\Phi\left(\frac{\alpha}{\sigma}\right),
$$
where $Y$ has the standard normal distribution and $\Phi$ is the cumulative distribution function of the standard normal distribution. Solving this equality, we obtain
\begin{equation}\label{alpha}
\alpha=\Phi^{-1}(1-\delta/2)\cdot \sigma.
\end{equation}

The usual magnitude of error in studies is $10^{-10}$ \cite{Jou12, Lev10, Jou13}, so let us fix $\delta/2=10^{-10}$; then $\Phi^{-1}(1-\delta/2)\approx 6.5$. That means, for example, in the case of the standard method
$$
\bE (T^{low})=T-6.5\sigma_1 
\quad \textrm{and} \quad
 \bE (V_\eps^{up})=V_\eps+6.5 s_1,
$$ 
with an error probability of $10^{-10}$. In real-life applications one should use their estimated values:
$$
\hat T^{low}=\hat T-6.5\hat\sigma_1 
\quad \textrm{and} \quad
\hat V_\eps^{up}=\hat V_\eps+6.5 \hat s_1.
$$


\subsection*{Appendix D: The maximal distance for CV QKD}


One can approximate the asymptotical key rate as an exponentially decreasing function of distance
(see Fig. \ref{fig_max_dist}). That is, we have
$$
K_\infty (T,V_\eps) \approx a \cdot 10^{-\kappa \cdot d}.
$$
We can obtain a trivial upper bound for the key rate:
$$
K < K_\infty (T,V_\eps) - \Delta(N).
$$ 
If the right-hand side drops below zero, the key rate will be negative. The right-hand side is positive if
$K_\infty (T,V_\eps) > \Delta(N)$, that is, if we use $\Delta(n)=\frac{c}{\sqrt{n}}$ \cite{Lev10} we have
$$
a \cdot 10^{-\kappa \cdot d} > \frac{c}{\sqrt{N}}.
$$
Rearranging this, we obtain a necessary condition for the positivity of the key rate:
\begin{equation}
d < \frac{1}{2\kappa} \cdot \log_{10} N -\frac{1}{\kappa} \log_{10}\frac{c}{a}.
\end{equation}
For realistic parameters $\kappa \approx 0.02$, so the coefficent of $\log_{10} N$ will be around 25, as is stated in the main text.


\subsection*{Appendix E: Dependence of key rate on parameters}

In the following we will show how the achievable distances for different methods change, if we vary the parameters of the protocol. We always optimize the modulation variance ($V$) and the ratio of states used for estimation ($r$). We check the key rate for coherent states ($V_S=1$), moderately squeezed states ($V_S=0.5$), and strongly squeezed states ($V_S=0.1$), while the distance is a function of $T$. So the only parameters remaining are $\eps$, $\beta$, and $N$.

\begin{figure}[!t]
\begin{center}
    \includegraphics[width=8.5cm]{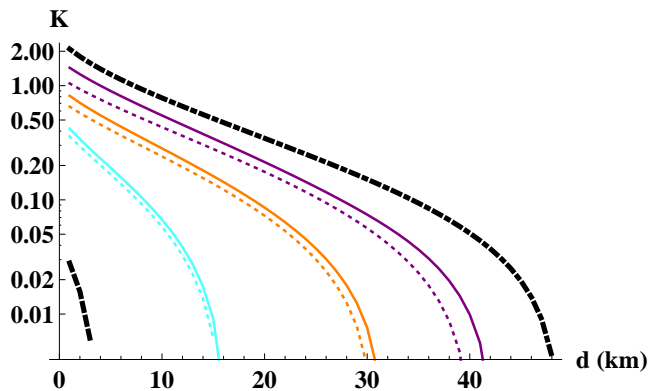}
 \caption{(Color online) The secure key rates of different methods: for coherent states [cyan (light gray), $V_S=1$], moderate squeezing [orange (medium gray), $V_S=0.5$], and strong squeezing [purple (dark gray), $V_S=0.1$], for the optimized single-modulation (dotted lines) and double-modulation (solid lines) schemes as a function of distance $d$ for $\eps=0.1$, $\beta=0.95$, and $N=10^6$. For comparison we plotted also the key rate yielded by the  current state-of-the-art technique (thick black dashed line) and the best theoretically possible upper limit $K^{th}$ (thick black dash-dotted line). \label{fig_eps}}
\end{center}
\end{figure}

If the excess noise $\eps$ increases  (see Fig. \ref{fig_eps}) the achievable distances become shorter in every case. But the difference for squeezed states will be much smaller than in the other cases (with the largest difference in the nonoptimized case). So we can conclude that the proposed optimized single- and double-modulation squeezed-state protocols are much more robust against noise.

\begin{figure}[!ht]
\begin{center}
   \includegraphics[width=8.5cm]{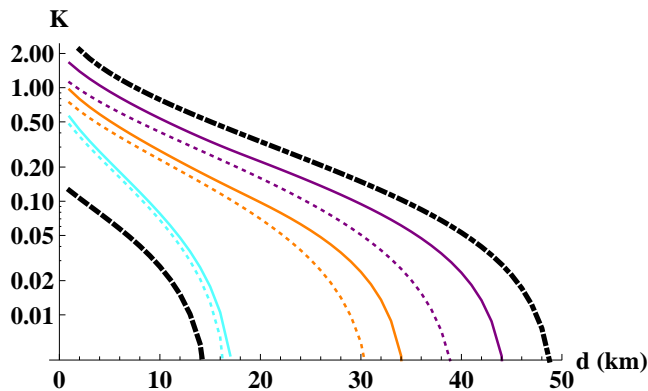}
\caption{(Color online)  The secure key rate of different methods: for coherent states [cyan (light gray), $V_S=1$], moderate squeezing [orange (medium gray), $V_S=0.5$], and strong squeezing [purple (dark gray), $V_S=0.1$], for the optimized single-modulation (dotted lines) and double-modulation  (solid lines) schemes as a function of distance $d$ for $\eps=0.01$, $\beta=0.8$, and $N=10^6$. For comparison we plotted also the key rate yielded by the  current state-of-the-art technique (thick black dashed line) and the best theoretically possible upper limit $K^{th}$ (thick black dash-dotted line).  \label{fig_beta}}
\end{center}
\end{figure}

The situation is similar if the reconciliation efficiency $\beta$ is reduced (see Fig. \ref{fig_beta}). Once again the distances are decreasing, but in this case the differences will be smaller. The nonoptimized version performs a little better than previously, because in this case the optimal $V$ becomes closer to the \textit{a priori} fixed modulation. The advantage of using the double-modulation method is more visible even for relatively large values of $T$ (i.e., for small distances).

\vfill\eject

\begin{figure}[!ht]
\begin{center}
   \includegraphics[width=8.5cm]{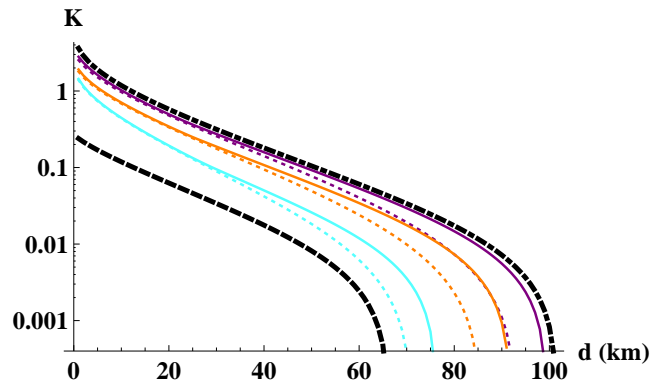}
\caption{(Color online) The secure key rates of different methods: for coherent states [cyan (light gray), $V_S=1$], moderate squeezing [orange (medium gray), $V_S=0.5$], and strong squeezing [purple (dark gray), $V_S=0.1$], for the optimized single-modulation (dotted lines) and double-modulation (solid lines) schemes as a function of distance $d$ for $\eps=0.01$, $\beta=0.95$, and $N=10^8$. For comparison we plotted also the key rate yielded by the  current state-of-the-art technique (thick black dashed line) and the best theoretically possible upper limit $K^{th}$ (thick black dash-dotted line). \label{fig7}}
\end{center}
\end{figure}

\noindent 
If we increase the block size $N$ (see Fig. \ref{fig7}) then the achievable distances will increase too. Note that the relation of the different methods is similar if $N$ is smaller; the improvement is close to an additive function (as we have seen for the theoretical limit in the main text). 
Let us also note that for large distances there will be a fair improvement using the double modulation method compared to the single modulation case: we can achieve the same distance with many fewer sqeezed states (e.g., double modulation with 3 dB squeezing produces a similar performance as single modulation with 10 dB squeezing). The key rate closely approaches the theoretical limit for the given level of squeezing.


\end{document}